\documentclass[journal]{IEEEtran}

\ifCLASSINFOpdf
\else
   \usepackage[dvips]{graphicx}
\fi
\usepackage{url}
\usepackage{graphicx}
\usepackage{comment}
\usepackage{amsmath,amssymb} 
\usepackage{color}
\usepackage{subfigure}
\usepackage{indentfirst}
\usepackage{stfloats}
\usepackage{epstopdf}
\usepackage{Chao_macros}
\begin{document}

\title{LEUGAN:Low-Light Image Enhancement by Unsupervised Generative Attentional Networks}

\author{Yangyang Qu, Chao liu and Yongsheng Ou
\thanks{This work was jointly supported by National Key Research and Development Program of China under Grant 2018AAA0103001 and National Natural Science Foundation of China (Grants No. U1613210)}
\thanks{Yangyang Qu, Chao Liu and Yongsheng Ou are with the Shenzhen Institutes of Advanced Technology, Chinese Academy of Sciences, Shenzhen 518055, China. (email: yy.qu@siat.ac.cn,chao.liu@siat.ac.cn,ys.ou@siat.ac.cn)}}

\markboth{Journal of \LaTeX\ Class Files, Vol. 14, No. 8, August 2015}
{Shell \MakeLowercase{\textit{et al.}}: Bare Demo of IEEEtran.cls for IEEE Journals}
\maketitle

\begin{abstract}
Restoring images from low-light data is a challenging problem. Most existing deep-network based algorithms are designed to be trained with pairwise images. Due to the lack of real-world datasets, they usually perform poorly when generalized in practice in terms of loss of image edge and color information. In this paper, we propose an unsupervised generation network with attention-guidance to handle the low-light image enhancement task. Specifically, our network contains two parts: an edge auxiliary module that restores sharper edges and an attention guidance module that recovers more realistic colors. Moreover, we propose a novel loss function to make the edges of the generated images more visible. Experiments validate that our proposed algorithm performs favorably against state-of-the-art methods, especially for real-world images in terms of image clarity and noise control.

\end{abstract}


\IEEEpeerreviewmaketitle

\section{Introduction}

\IEEEPARstart{I}{n} recent years, due to the rapid popularity of cameras equipped in various devices, image acquisition and vision algorithms are more and more widely used. While the images taken under low-light conditions may cause problems, such as low contrast, blurred edges, poor color, and noise further producing negative impacts in the depth estimation and semantic segmentation. Raising ISO is capable of increasing the image sensor sensitivity, but it will also amplify the noise in the meantime.
\par Histogram equalization (HE) \cite{2} is one of the most widely adopted techniques to enhance low-light images because of its simplicity and effectiveness. However, there are some shortcomings, such as excessive contrast adjustment, noise amplification, and contour distortion. Various studies have been carried out to address these disadvantages. Wang et al. \cite{LDR} proposed a novel contrast enhancement algorithm based on the layered difference representation of 2D histograms. Another algorithm \cite{CVC} enhances the contrast of an input image using interpixel contextual information.
\par Retinex theory \cite{3} is another popular low-light enhancement technique, which assumes that images can be decomposed into reflectance and illumination. Wanget al. \cite{Wang} presents a bright-pass filter to decompose the observed image and then preserves the naturalness while enhancing the image details. In \cite{LIME}, the brightness of each pixel is replaced by the maximum value in R, G, and B channels to generate an enhanced map.
\par Learning-based methods have been studied extensively in the past few years.  Gharbi et al. \cite{Gharbi} consider a convolutional neural network that directly learns an end-to-end mapping between dark and bright images. Chen et al. \cite{GAN} developed an unpaired learning model for photo enhancement based on a two-way generative adversarial network (GAN).  Lv et al. \cite{MBLLEN} propose a multi-branch low-light enhancement network via multiple subnets and multi-branch fusion. Wang et al.\cite{GLAD} calculates a global illumination estimation for the input and adjust the illumination under the guidance of the estimation. Though these methods achieve significant progress, enhancing low-light images is still challenging. The most significant bottleneck is the requirement for real-world pairwise images which severely limits the usage in practice.
\par To solve these issues, we propose a novel unsupervised generative attentional network for low-light enhancement. The contributions of this work  are summarized as follows:
\begin{figure}[t]
\begin{center}
\begin{tabular}{cccc}
\includegraphics[width = 0.21\linewidth]{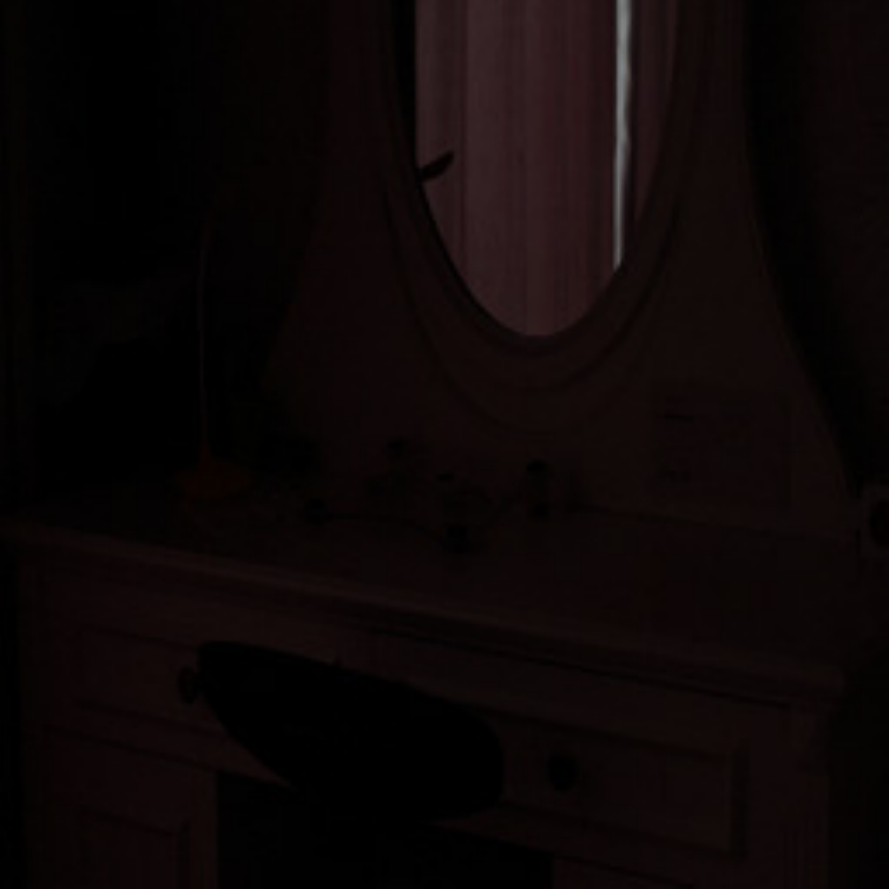}&\hspace{-3mm}
\includegraphics[width = 0.21\linewidth]{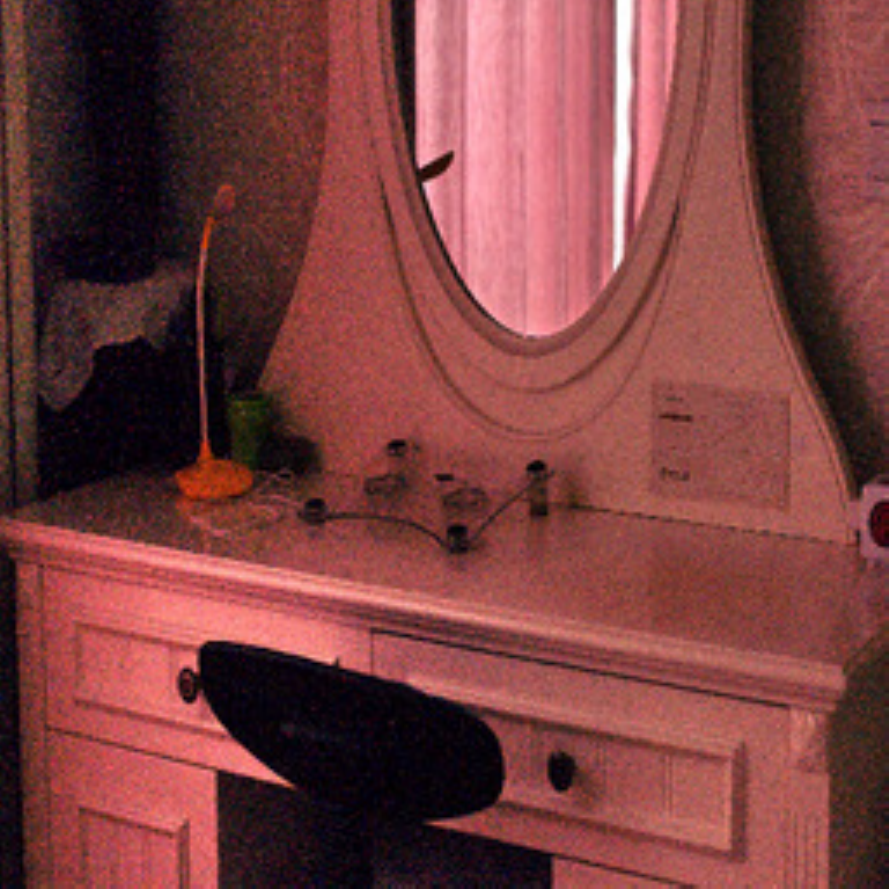}&\hspace{-3mm}
\includegraphics[width =0.21\linewidth]{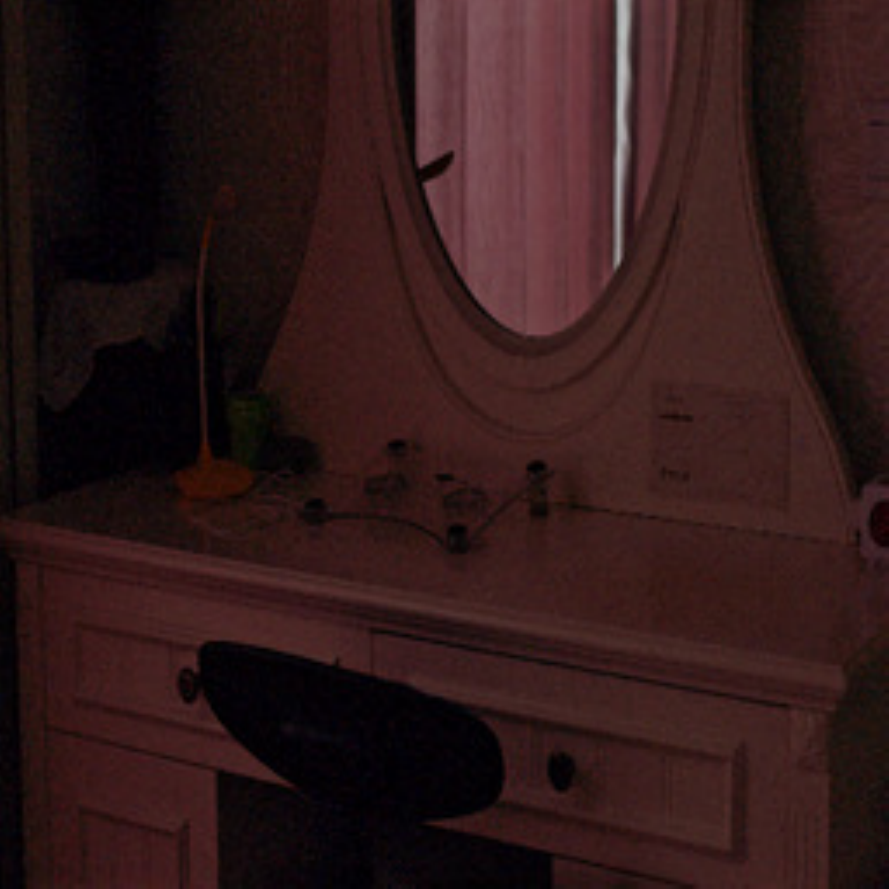}&\hspace{-3mm}
\includegraphics[width =0.21\linewidth]{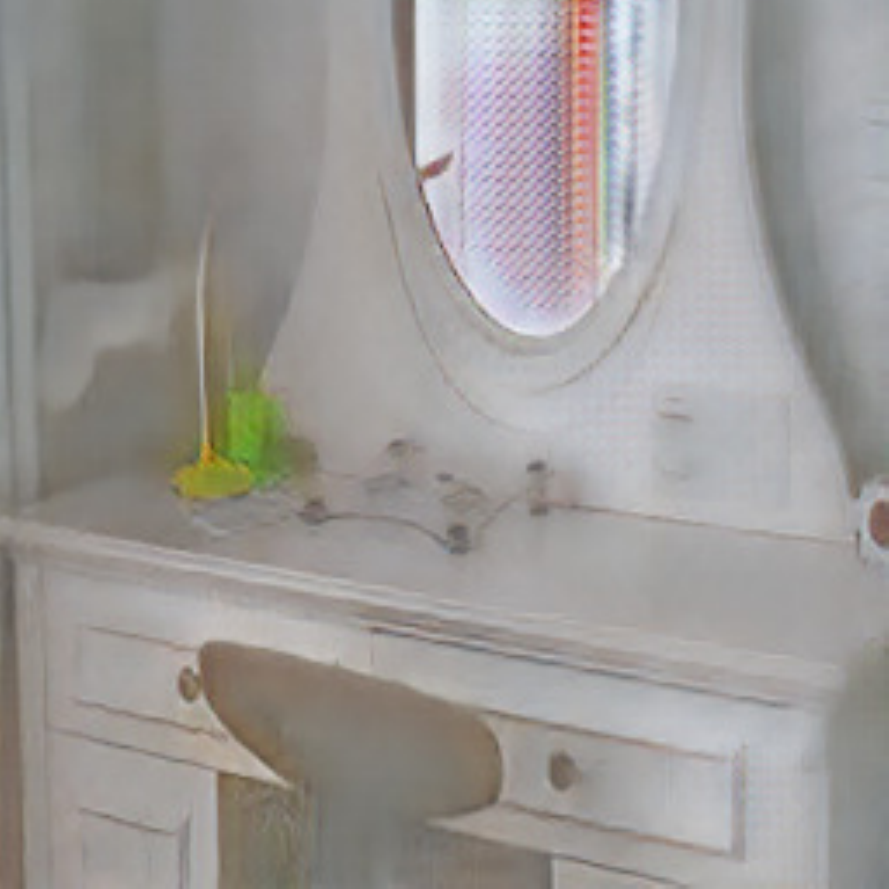}
\\
Input &\hspace{-3mm} LIME  &\hspace{-3mm}  SRIE &\hspace{-3mm} CycleGAN
\\
\includegraphics[width = 0.21\linewidth]{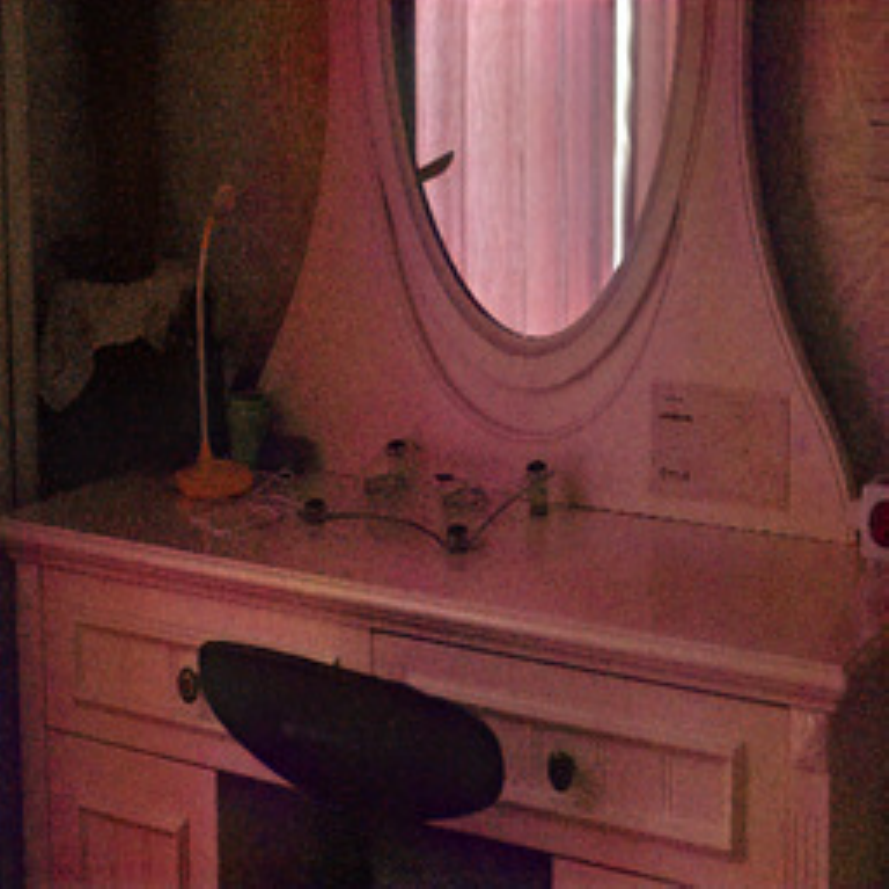}&\hspace{-3mm}
\includegraphics[width =0.21\linewidth]{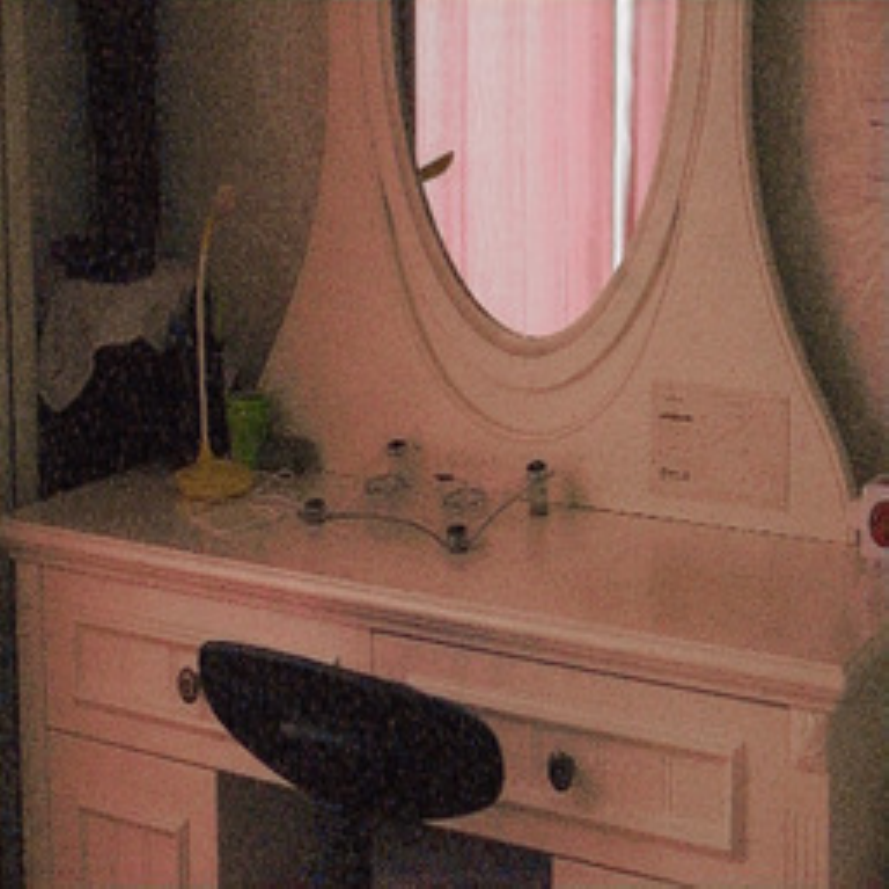}&\hspace{-3mm}
\includegraphics[width =0.21\linewidth]{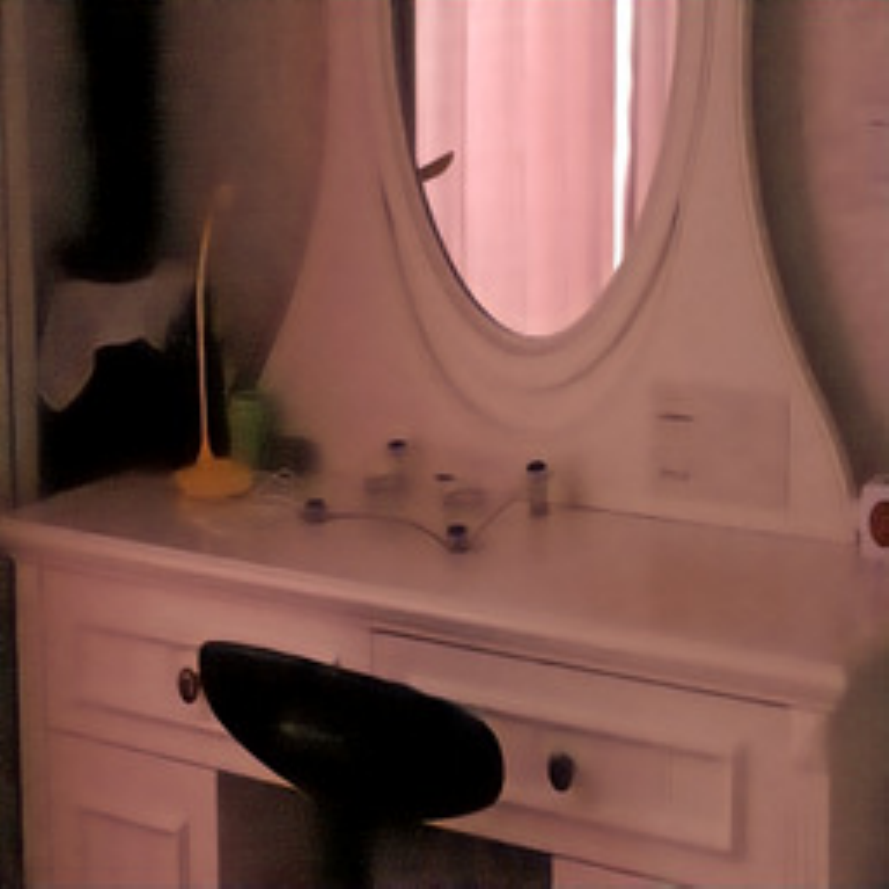}&\hspace{-3mm}
\includegraphics[width = 0.21\linewidth]{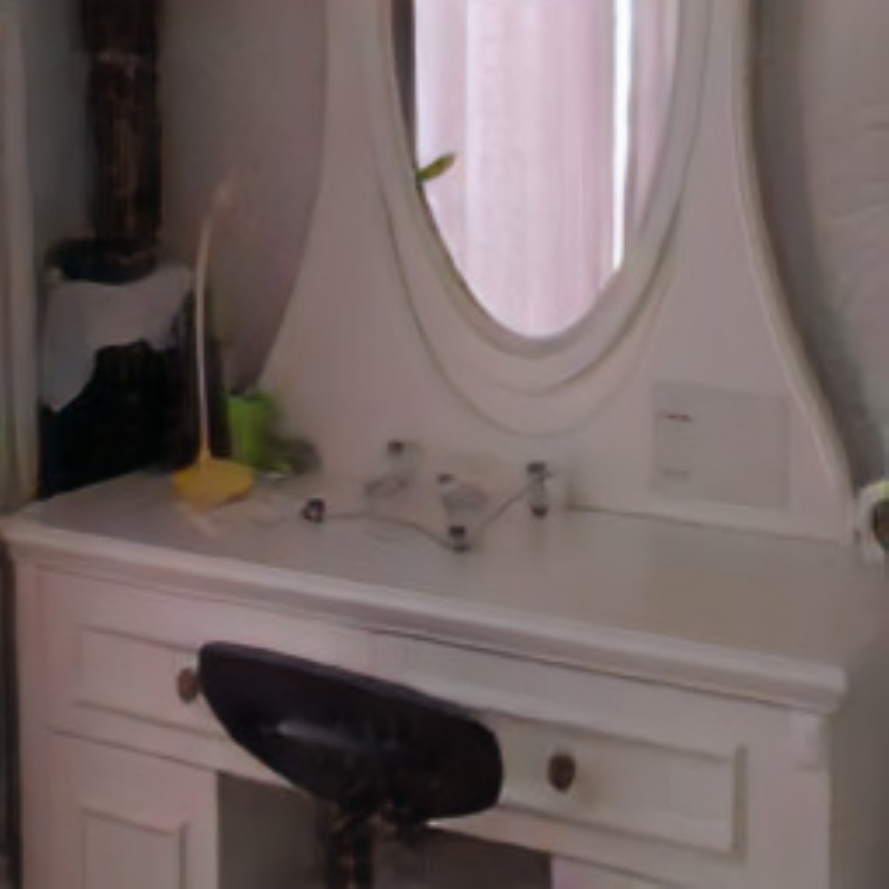}
\\
EnlightenGAN &\hspace{-3mm} GLAD  &\hspace{-3mm}  MBLLEN &\hspace{-3mm} LEUGAN
\\
\end{tabular}
\end{center}
\caption{Comparison with other state-of-the-art methods.  As can be seen, there are still noise residuals in the enhanced images of existing methods. LEUGAN successfully suppresses the noise, gets the most normal color and produces the best visible details.}
\label{fig:2}
\end{figure}

\begin{itemize}
\item We develop an unsupervised generative attentional network that can be trained with unpaired image sets.
\item We design an edge enhancement module that recovers sharper edges and an attention module that improves the color and texture restoring at focused areas.
\item The experimental results demonstrate the effectiveness of the proposed network in terms of image clarity and noise control.
\end{itemize}

\begin{figure*}[t]
\centering
\includegraphics[height=8cm]{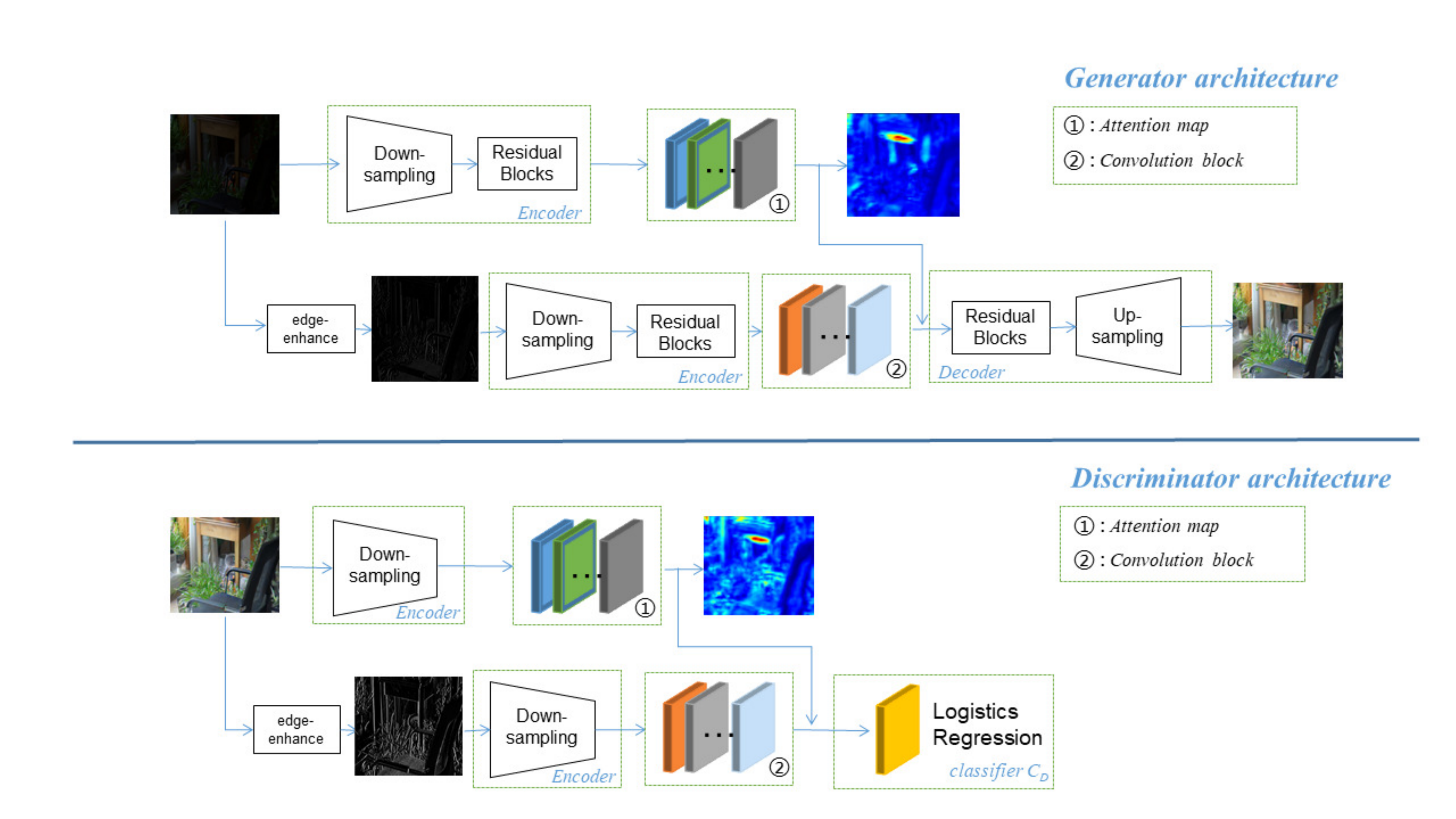}
\caption{The architecture of the proposed low-light enhancement network. The proposed LEUGAN consists of two branches, one is to process the original image, and the other is to enhance the original image after edge detection. Specific details are described in Sec.~\ref{sec: Network Architecture}.}
\label{fig:moxing}
\end{figure*}

\section{Low-Light Image Enhancement by Unsupervised Generative Attentional Networks}
In unsupervised image-to-image translation, a joint distribution is yielded through  models where the network encodes images from  two domains into a shared feature space.
Our goal is to map from a source domain $\alpha$ (defined by low-light images) to a target domain $\beta$ (defined by normal-light images) only by unpaired samples via a network model that consists of two generators and two discriminators, as shown in Fig.~\ref{fig:moxing}.
\par We integrate an edge-enhanced module and an attention module into the networks.
The edge module focus on recognizing edges and generates clearer textures, while the pixel and channel attention module leads the networks to mark particular areas for restoring richer details and better colors.

\subsection{Network Architecture of Generative Model}\label{sec: Network Architecture}
As shown in Fig.~\ref{fig:moxing}, our generator model $G_{\alpha\rightarrow\beta}$ consists of an edge-enhancement module $ED_{G}$, an encoder $E_{G}$, a decoder $D_{G}$, and an attention module  $U_{G}$ which focuses on particular regions.


Our generator contains two branches. For the upper branch, our purpose is to obtain the attention map $\MI_{att}$ of the image. Firstly, the original low-light image $\MI_{low}$ is down sampled by encoder module $E_{G}$. Then we input the encoder result $\Ve_{r}$ into attention module $A_{G}$ to calculate the important distribution of features. For another branch, we input $\MI_{low}$ to the edge module $ED_{G}$ to get the edge image $\MI_{edg}$. Then we use the same encoder module $E_{G}$ for down sampling. We calculate the attention map $\MI_{att}$ from the result of down sampling, and get the final normal-illumination image $\MI_{nor}$ via decoder $D_{G}$.


\subsubsection{Attention Module}
\par Considering the uneven distribution of light, the pixel attention would be more attentive to the informative areas, such as the the texture-rich regions. For the pixel attention, we calculate the feature map of each channel after the pixel attention. The typical attention modules used in other networks handle images on only pixel level without taking advantage of interrelation between color channels. Therefore, we add a channel attention part inspired by the Squeeze-and-Excitation networks  \cite{51}, which aims to adaptively recalibrate channel-wise feature
responses.
\par Firstly, we calculate an auxiliary feature $\mathbf{u} \in \mathbb{R}^{H \times W\times C}$, from the encoded result, $\ME_{r} \in \mathbb{R}^{H^{\prime} \times W^{\prime} \times C^{\prime}}$. By using a three-dimension convolution kernel $\mathbf{v}$, we can get the convolution output ${\mathbf{u} =\mathbf{v} * \ME_{r}}$. 
\par After that, we calculate the pixel attention. After getting the important distribution through global average pooling $\Vf_{ave}$ and maximum pooling $\Vf_{max}$, we can get the feature map ${\Vf_{G}=[\Vf_{ave}^{\T},\Vf_{max}^{\T}]^{\T}}.$
 And then we can get the standard attention map $\MI_{att} $:
\begin{equation}
\MI_{att} =\Vf_{G} * \mathbf{u},
\end{equation}
and the visualization of  attention map can be seen in Fig.~\ref{fig:moxing}.
\par In order to make our model more flexible to adjust the attention map and improve the training effect of the model, we add residual connection after the attention map which is inspired by \cite{NICE}. Assuming that the result of our encoder is $\ME_r$, we can get the dynamic $\ME_{ro}$:
\begin{equation}
\ME_{ro} =\lambda * \MI_{att} * \ME_r + \ME_r,
\end{equation}
 where $\lambda$ is a trainable parameter that dynamically adjusts to the relationship between the feature map $\MI_{att}$ and the original result $\ME_r$, e.g.  When $\lambda$=0, results from attention module are discarded, and the encoder output is directly passed to the decoder.
\subsubsection{Edge Module}
To extract edge information from low-light images, we first convert the input to gray-scale image, and then  we get the oriented gradient via convolution with a improved sobel operator for for horizontal direction and vertical direction respectively. Being different from \cite{29}, we add edge module in the process of generator and discriminator.

\subsection{Network Architecture of Discriminator Model}
Similar to the generator models, our discriminator model $D_{\alpha\rightarrow\beta}$ consists of edge-enhancement module, an encoder $E_{D}$, a decoder $D_{D}$, an attention module $A_{D}$ and a classifier $C_{D}$. With the help of these modules, the model will discriminate whether $x$ comes from the target domain or the translated source domain $G_{\alpha\rightarrow\beta}$. When we get an image $\Vx$, $D_{\alpha\rightarrow\beta}(x)$ uses attention map to get:
\begin{equation}
\Vc_{r}(x)=\Ve_{D} * \MI_{att,D}(x),
\end{equation}
 where  $\Ve_{D}$ is the meaning of encoder result, $ I_{attD}(x)$ is the attention maps which is trained by $A_{D}$, and $\Vc_{r}$ is the classifier result.
\subsection{Loss function}
The loss function comprises five loss terms which contains a new loss function proposed by us.
\subsubsection{Structural Loss.}
We propose a loss function by further considering the structural information of the image. $\mu_{x}$ and $\mu_{y}$ represents the pixel value averages of the image, $\lambda_{x}$ means the standard deviation of the image,and $\lambda_{y}$ has the same meaning as $\lambda_{x}$.  $\theta$ is a constant and we set it to 1. We formulate the loss functions as:
\begin{align}
&\lambda_{x}=\sqrt{\left(\frac{1}{N-1} \sum_{i=1}^{N}\left(x_{i}-\mu_{x}\right)^{2}\right)},\\
&\lambda_{x y}=\frac{1}{N-1} \sum_{i=1}^{N}\left(x_{i}-\mu_{x}\right)\left(y_{i}-\mu_{y}\right),\\
&{\mathcal{L}_{str}}=\frac{\lambda_{x y}+ \theta}{\lambda_{x} \lambda_{y}+\theta},
\end{align}
where N is the number of pixels in the image.
\subsubsection{Identity Loss.}
After considering the structural information, we use identity loss for the generator to ensure that the color distributions of input image $\mathit{x}$ which is from $\beta$ domain (normal-light image) and output image generated by $G_{\alpha\rightarrow\beta}$ is similar. The loss function is given by:
\begin{align}
\mathcal{L}_{idt}=\mathbb{E}_{x \sim X_{\beta}}\left[\left|x-G_{\alpha\rightarrow\beta}(x)\right|\right],
\end{align}
where$|\cdot|$ indicates absolute value.
\subsubsection{Adversarial Loss}
We adopt the adversarial loss to minimize the distance between the real and output normal light distributions.
\begin{equation}
\begin{aligned}
\mathcal{L}_{adv} &=\mathbb{E}_{x_{\beta} \sim X_{\beta}}\left[\log (D_{\beta}(x)\right)] \\
&+\mathbb{E}_{x_{\alpha} \sim X_{\alpha}}\left[\log \left(1-D_{\beta}(G_{\alpha\rightarrow\beta}(x)\right)\right].
\end{aligned}
\end{equation}
\subsubsection{Cycle Consistency Loss}
This loss is proposed to tackle the unpaired translation problem  \cite{46}, and it has been used wildly to prevent the network from generating random images in the target domain. We also put it in our network, and formulate the objective loss functions as:
\begin{equation}
\begin{aligned}
\mathcal{L}_{cyc} &=\mathbb{E}_{x \sim X_{\alpha}}\left[|x-G_{\alpha\rightarrow\beta}(G_{\beta\rightarrow\alpha}(x))|\right].
\end{aligned}
\end{equation}
\subsubsection{Auxiliary Loss.}
Here we adopt the auxiliary loss to get to know where is the most difference between two domains in the current state, $\eta_{D_{\beta}}$ represents the auxiliary classifiers, and it is defined as:
\begin{equation}
\begin{aligned}
\mathcal{L}_{aux} &=\mathbb{E}_{x \sim X_{\beta}}\left[\left(\eta_{D_{\beta}}(x)\right)^{2}\right] \\
&+\mathbb{E}_{x \sim X_{\alpha}}\left[2*\log \left(1-\eta_{D_{\beta}}\left(G_{\alpha\rightarrow\beta}(x)\right)\right)\right],
\end{aligned}
\end{equation}
where $\eta_{D_{\beta}}$ is calculated as:
\begin{equation}
\eta_{s}(x)=\sigma\left(\sum^{k} w^{k} \sum E_{r}^{k}(x)\right),
\end{equation}
where $\sigma(\cdot)$ is softmax function \cite{50}.
\par The overall loss function for training LEUGAN is written as follows:
\begin{multline}
\mathcal{L}_{all}=\omega_{s t} \mathcal{L}_{s t r}+\omega_{s i} \mathcal{L}_{i d t} \\
+\omega_{s d} \mathcal{L}_{a d v}+\omega_{s y} \mathcal{L}_{c y c}+\omega_{s u} \mathcal{L}_{a u x}.
\end{multline}
where ${\mathcal{L}_{str}}$, $\mathcal{L}_{idt}$, $\mathcal{L}_{adv}$, $\mathcal{L}_{cyc}$, $\mathcal{L}_{aux}$ represent the corresponding loss function, and $\omega_{s t}$, $\omega_{s i}$, $\omega_{s d}$, $\omega_{s y}$, $\omega_{s u}$ are the corresponding coefficients of these losses. In the experiments, we empirically set $\omega_{s t}$, = 1, $\omega_{s i}$ =10, $\omega_{s d}$=10, $\omega_{s y}$=10, $\omega_{s u}$=100.
\begin{figure*}[t]
\begin{center}
\begin{tabular}{cccccccc}
\includegraphics[width = 0.12\linewidth]{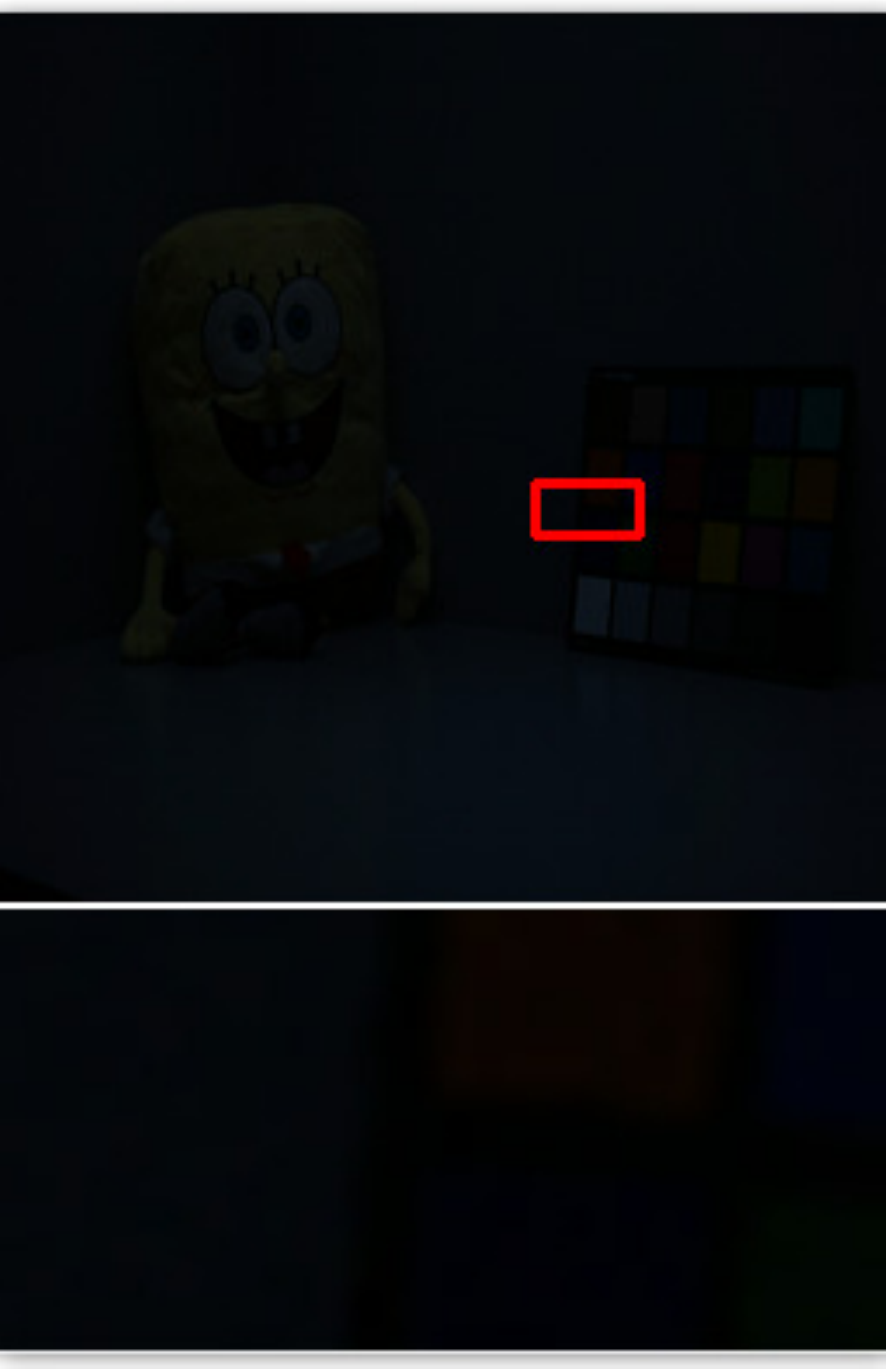} &\hspace{-5mm}
\includegraphics[width = 0.12\linewidth]{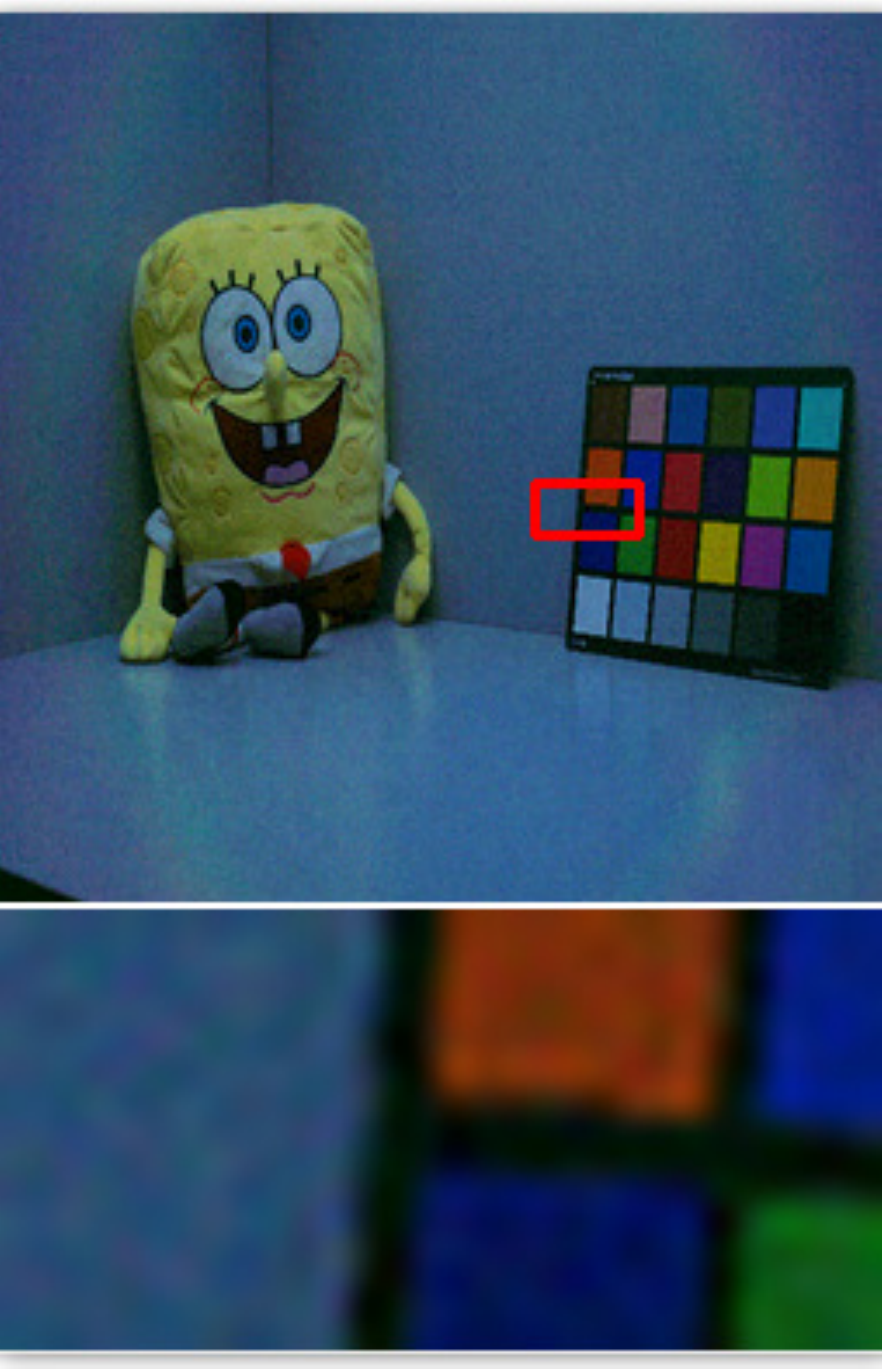}&\hspace{-5mm}
\includegraphics[width = 0.12\linewidth]{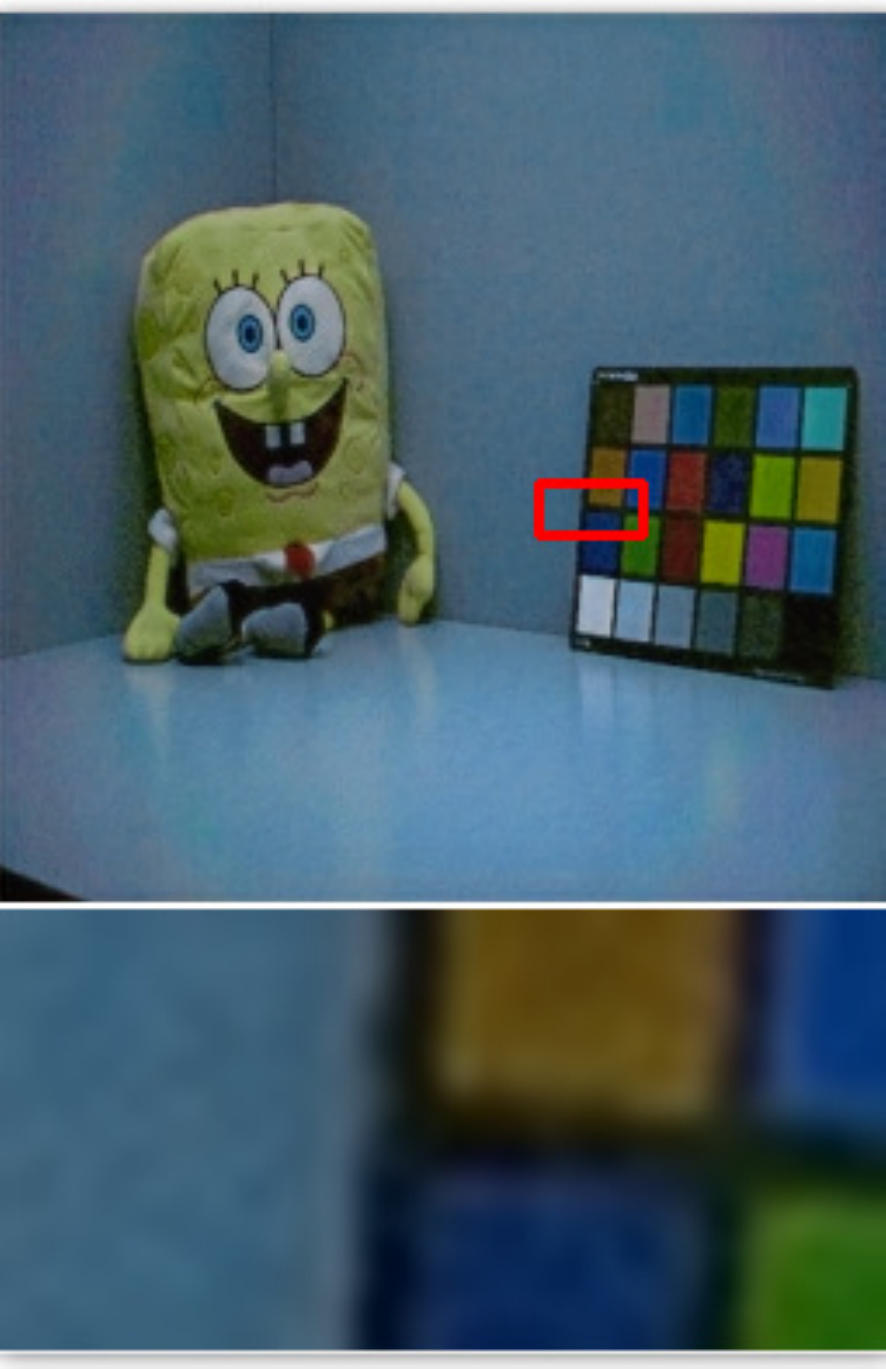}&\hspace{-5mm}
\includegraphics[width = 0.12\linewidth]{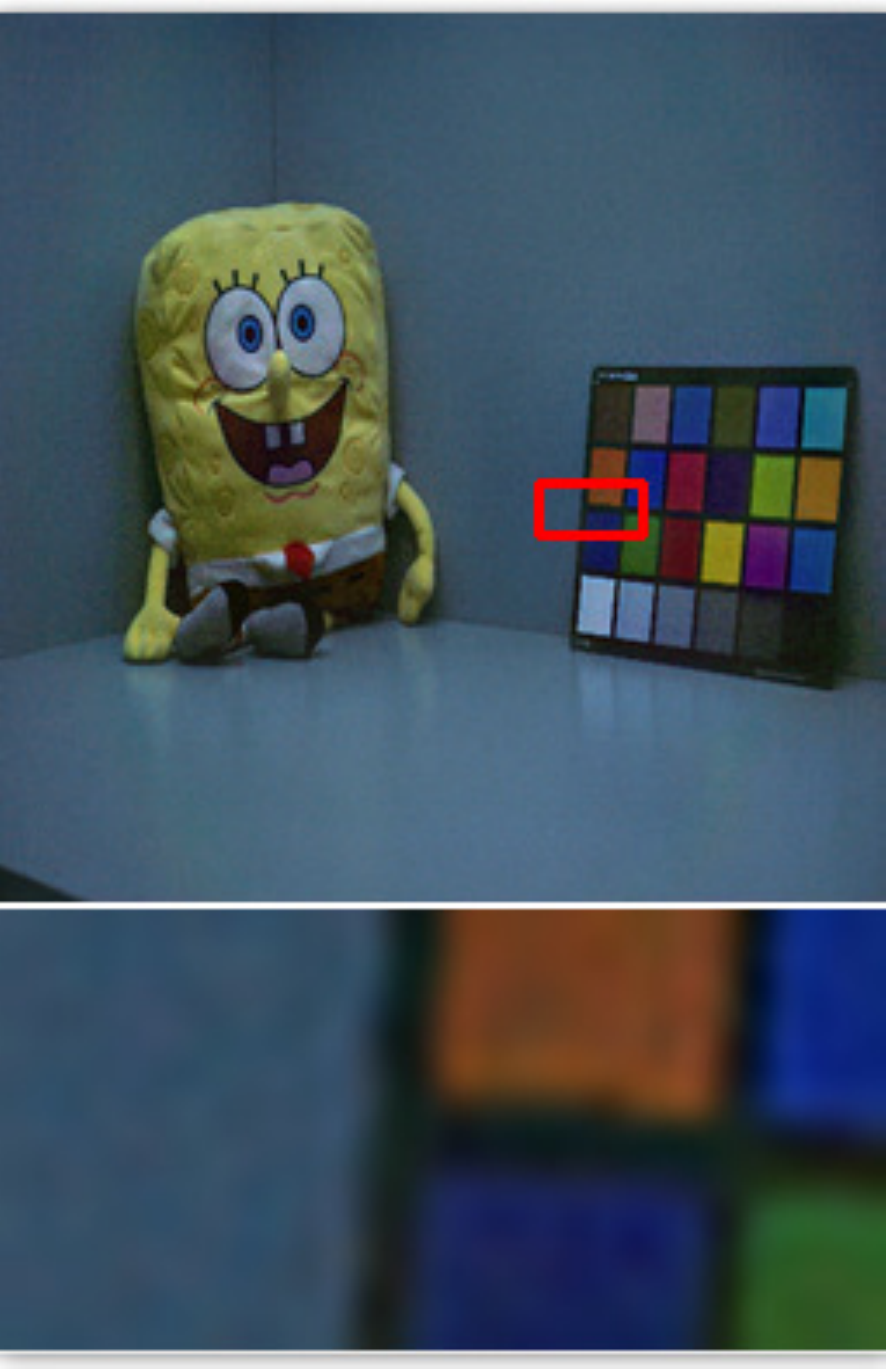}&\hspace{-5mm}
\includegraphics[width = 0.12\linewidth]{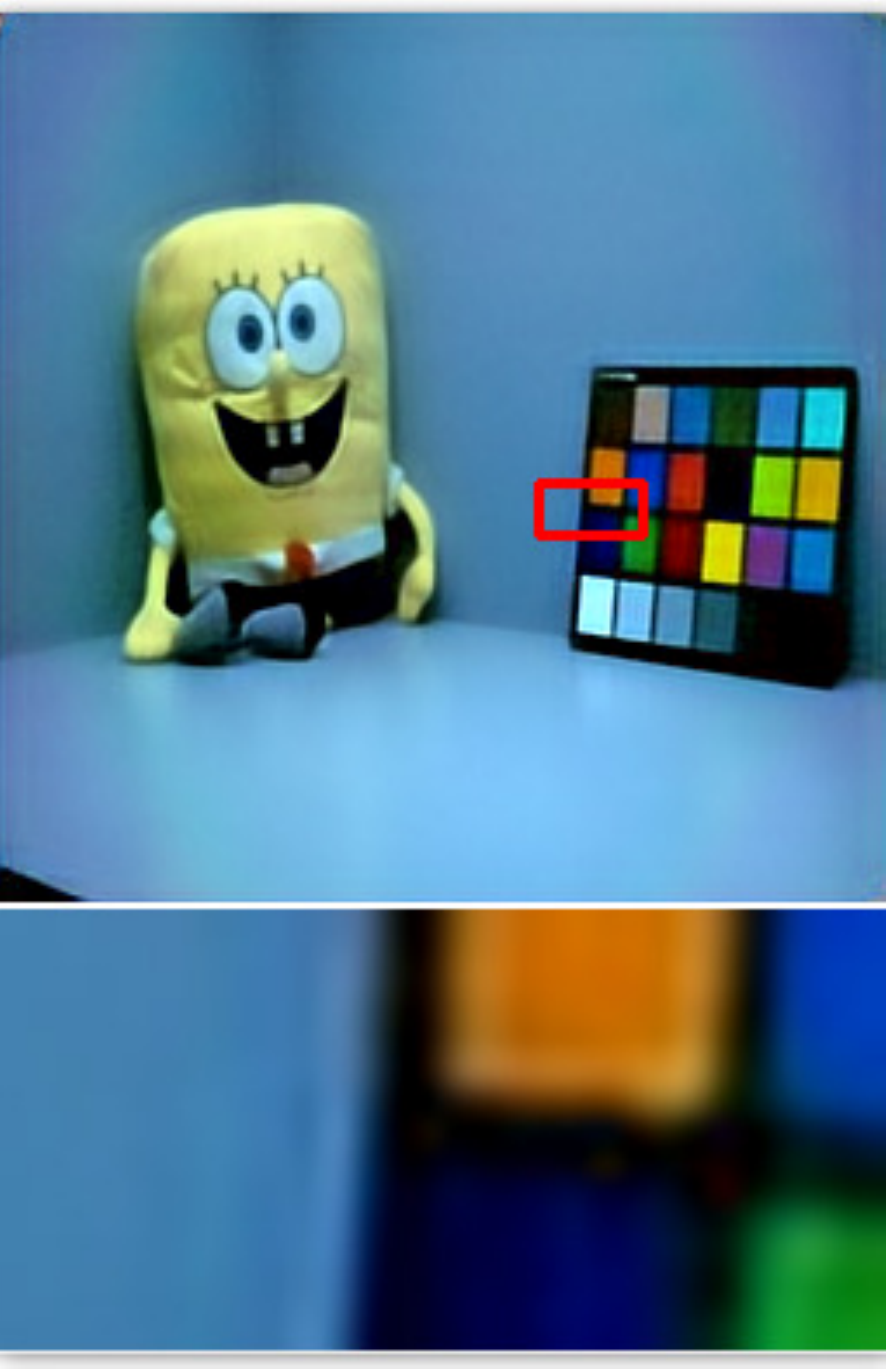}&\hspace{-5mm}
\includegraphics[width = 0.12\linewidth]{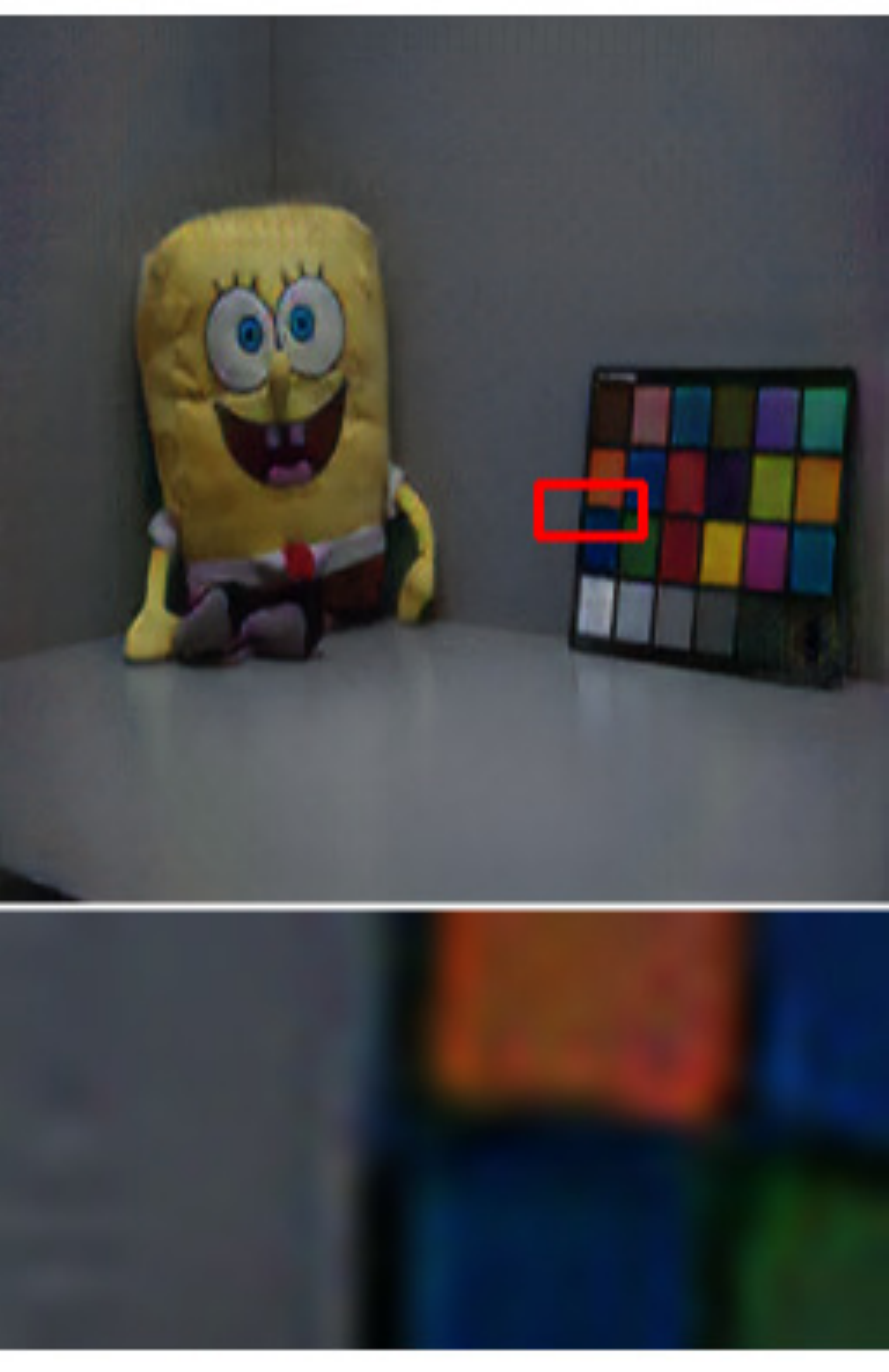}&\hspace{-5mm}
\includegraphics[width = 0.12\linewidth]{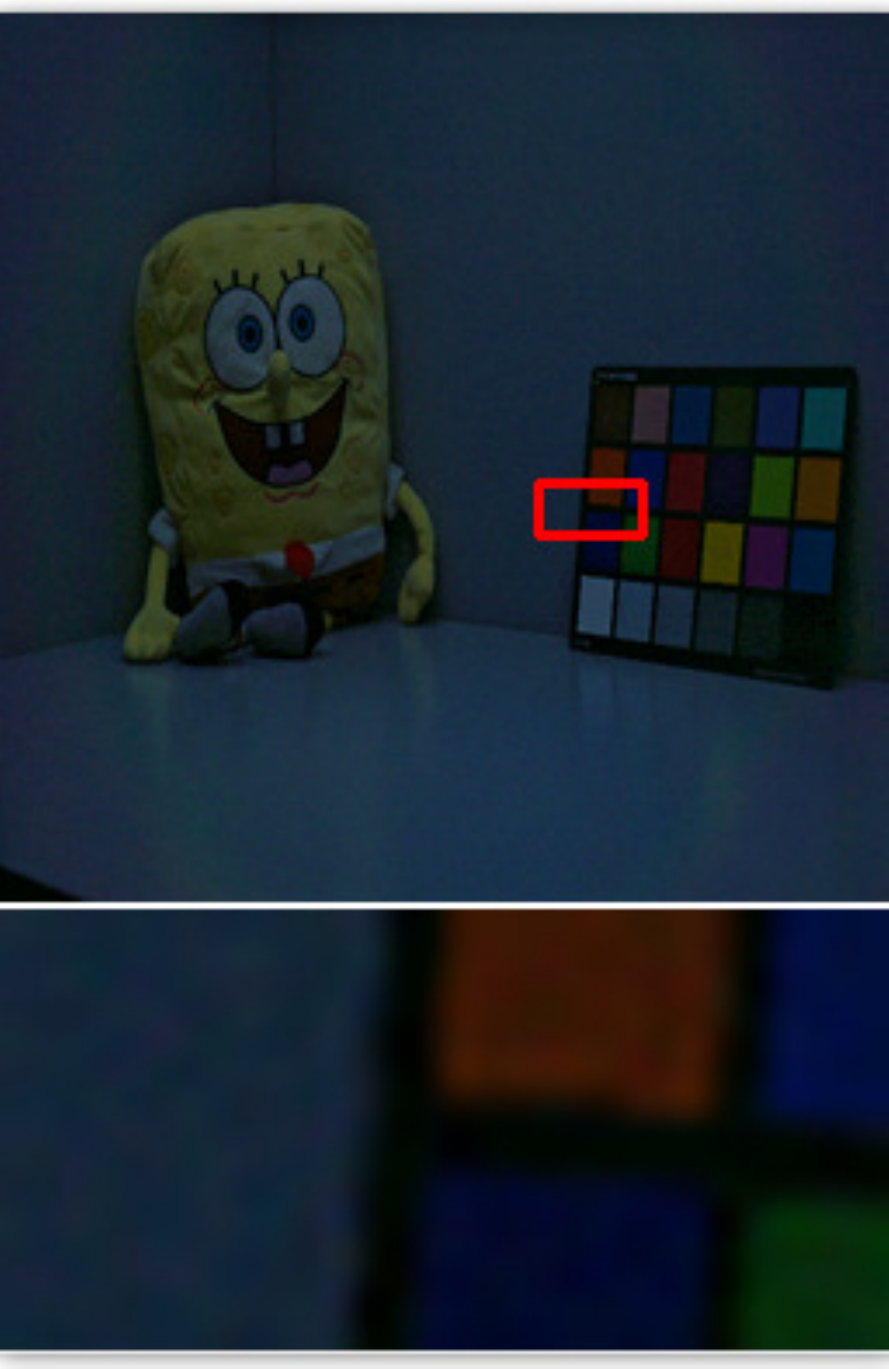}&\hspace{-5mm}
\includegraphics[width = 0.12\linewidth]{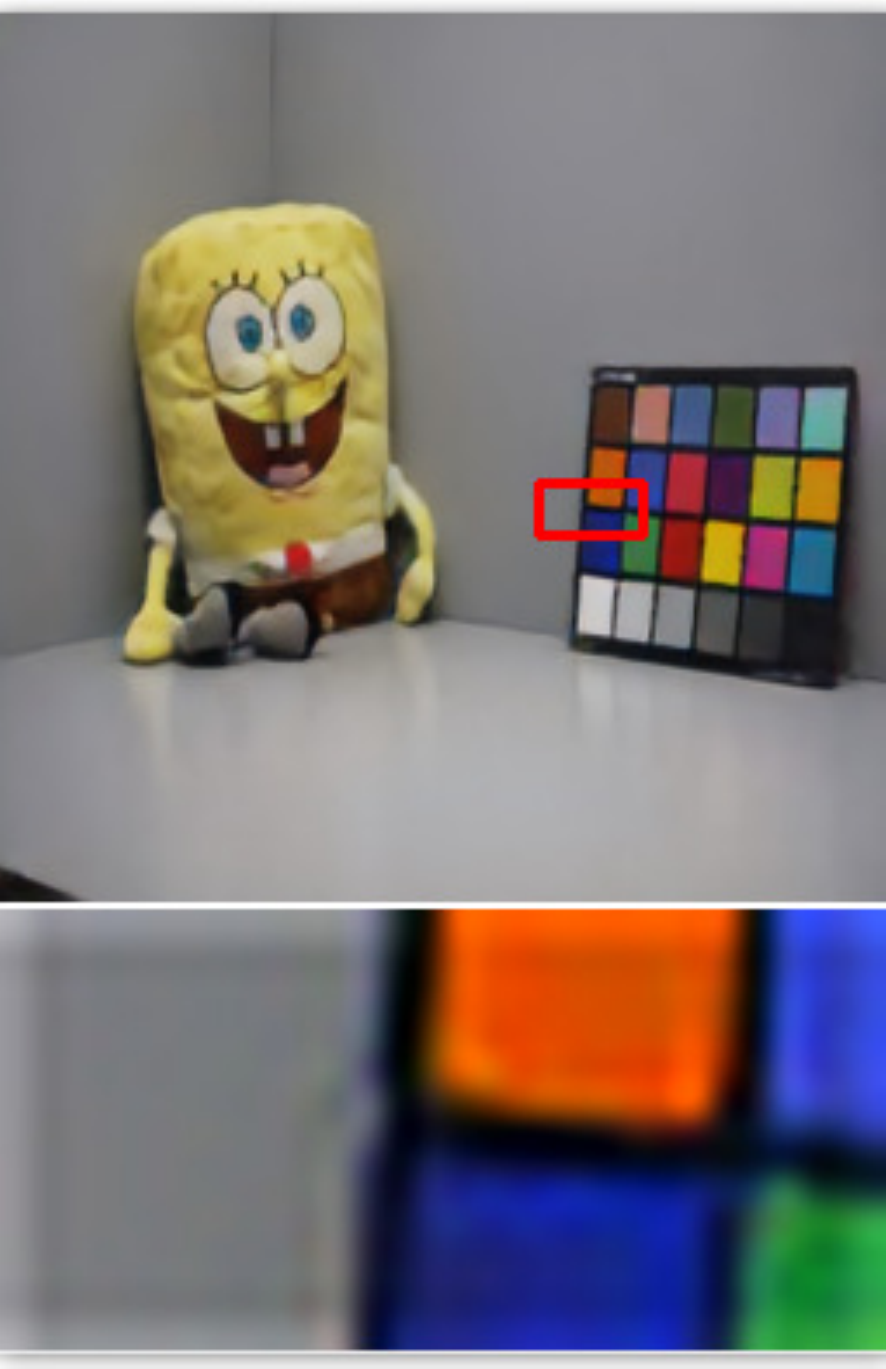}
\\
Input &\hspace{-5mm} LIME  &\hspace{-5mm}  GLAD &\hspace{-5mm} EnlightenGAN &\hspace{-5mm} MBLLEN &\hspace{-5mm} CycleGAN &\hspace{-5mm} SRIE&\hspace{-5mm} LEUGAN
\\
\end{tabular}
\end{center}
\caption{ Visual comparison with state-of-the-art low-light image enhancement methods. Zoom-in regions are used to illustrate the visual difference. It can be seen  that our results contain little noise and obtain the most natural color of the image.}
\label{fig:duibi}
\end{figure*}

\section{Experiments}
To validate LEUGAN, we evaluate the performance on Low-Light(LOL) dataset \cite{57}, and we compare it with LIME \cite{10}, SRIE \cite{7}, CycleGAN \cite{cyclegan}, EnlightenGAN \cite{31}, MBLLEN \cite{MBLLEN} and GLADNet \cite{GLAD} which are all trained on the LOL dataset.
\subsection{Implementation Details}
The proposed network is trained using the proposed dataset. The parameters of the networks were optimized using the Adabound algorithm \cite{57}. As for the normalization method, we adopt the AdaLIN [12] whose parameters are dynamically computed from the attention map, and we train the model with a weight decay learning rate 0.0001.
\par In the generator, we use the instance normalization for the encoder to increase the accuracy of the auxiliary attention. We only use AdaLIN for the decoder. In the discriminator, we use spectral normalization, and we employ two different scales of PatchGAN \cite{15} which classifies whether  local (70 $\times$ 70) and global (256 $\times$ 256) image patches are real or fake. After that, we use ReLU \cite{relu} in the generator and leaky-ReLU with a slope of 0.2 for the activation function.
\subsection{No-Reference Quality Assessment}
The sharpness and the amount of noise are important indicators for measuring the quality of the image, and it can better correspond to human subjective feelings.
For the non-reference image quality evaluation, we use several representative definition algorithms for discussion and analysis.
\par We adopt natural image quality evaluator (NIQE) \cite{NIQE} which is a no-reference image quality evaluation standard to justify how "naturally" the image looks like. It has better consistency with human subjective quality evaluation. Lower the NIQE score, more natural the image looks. Besides, we also use Vollath \cite{Vollath} to measure image sharpness, PCA-based \cite{60} to measure the image noise. Higher Vollath means better clarity, and lower PCA-based indicates less image noise.

Tab.~\ref{table:1} reports the quantitative results of the low-light image enhancement. Bold blue represents the best result, and bold black represents the second best result. The results show that our superiority in visual effects and noise control. From the results we can find that our method achieves the best results in terms of visual effect, and it is much better than other methods in terms of noise control. We also achieved the second best result in terms of clarity. We also confirmed our experimental results in the latter visual effect analysis.
\subsection{Visual effect analysis}
We give an example to illustrate the superiority of our method in Fig.~\ref{fig:duibi}, the last column is our method.
\par Through the results shown in Fig.~\ref{fig:duibi}, we can find that our method achieves good results in both color restoration and noise control.  EnlightenGAN \cite{31} which is an unsupervised method has achieved good results in terms of noise control, while it does not perform well in color. CycleGAN \cite{cyclegan}  generates darker result compared with our method. The results of other methods are also unsatisfactory in terms of color and brightness. As can be seen that our method restores more realistic image with clearer details, better contrast and normal brightness.

\begin{figure}[t]
\begin{center}
\begin{tabular}{ccccc}
\includegraphics[width = 0.19\linewidth]{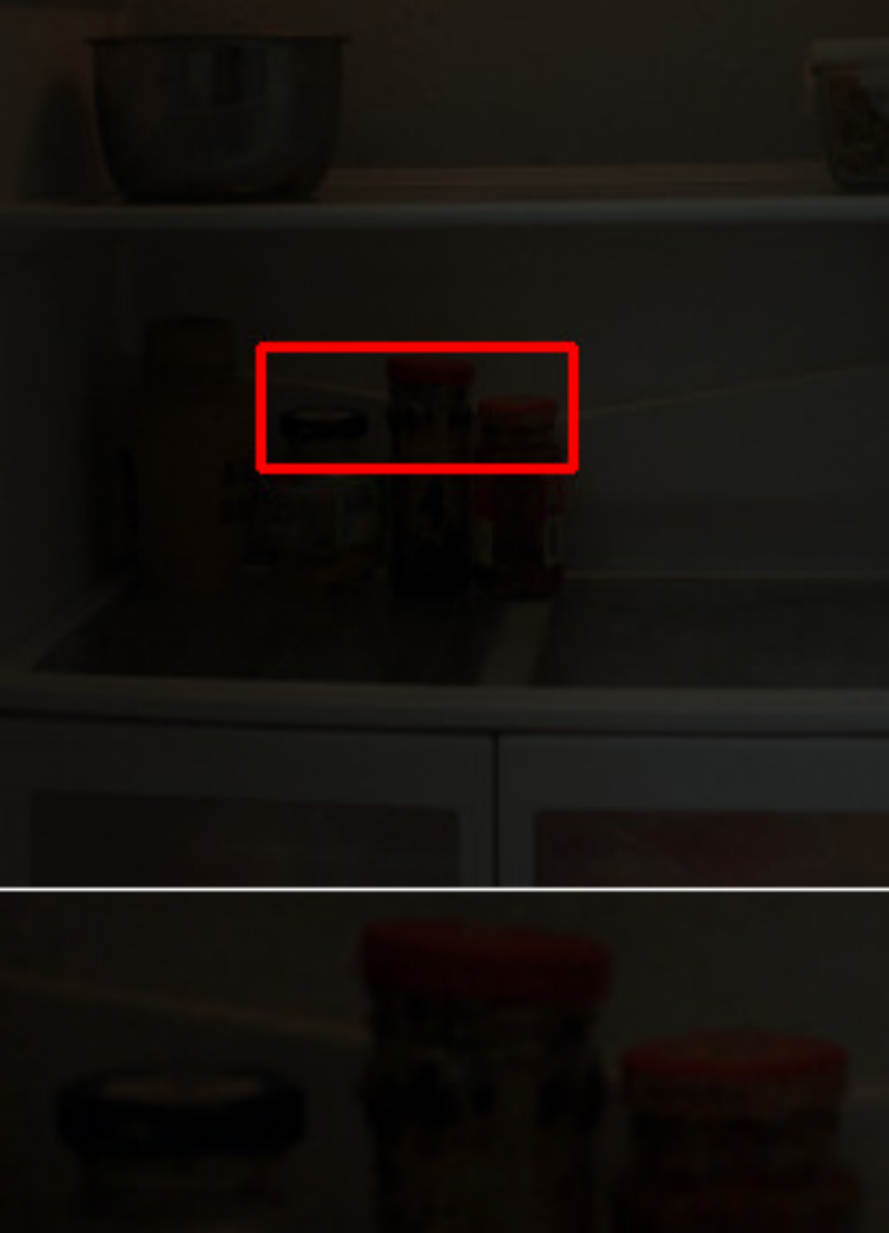} &\hspace{-5mm}
\includegraphics[width = 0.19\linewidth]{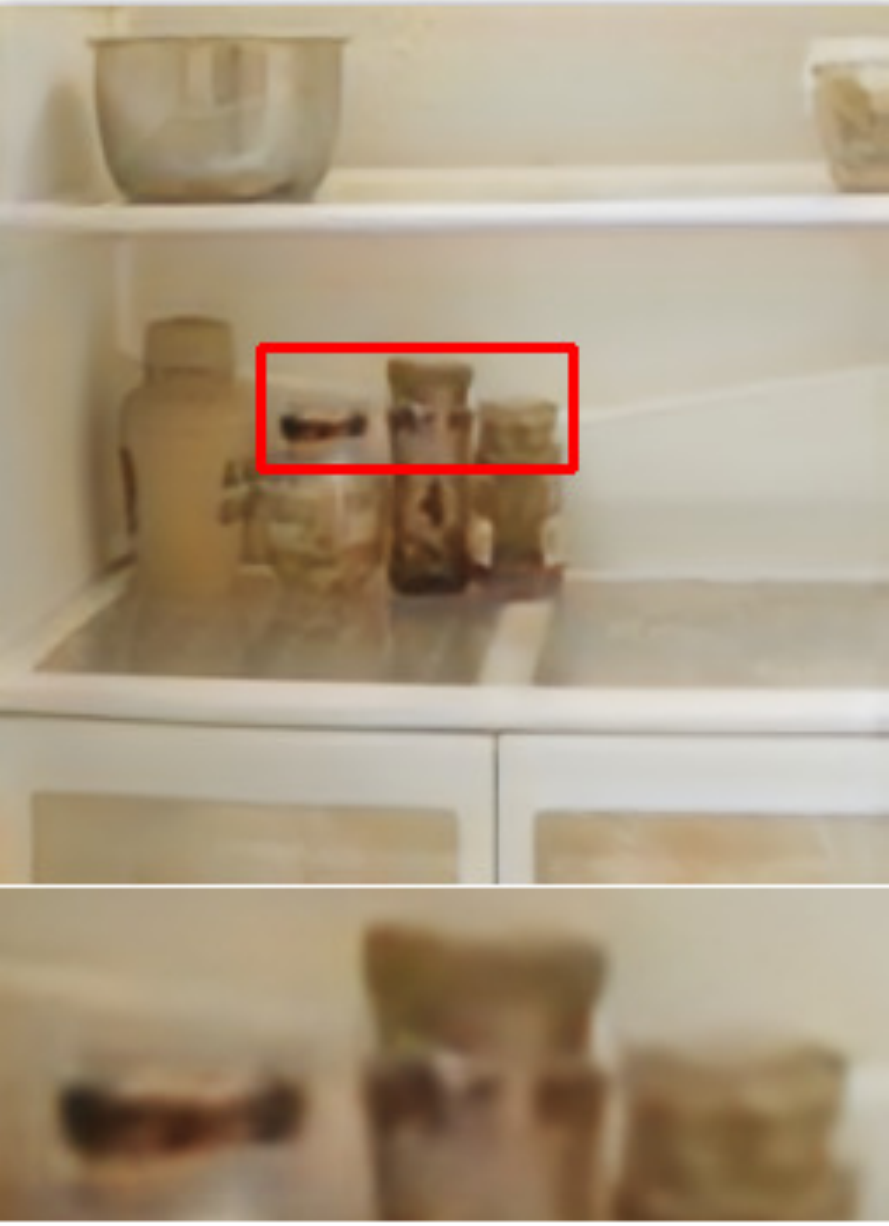} &\hspace{-5mm}
\includegraphics[width = 0.19\linewidth]{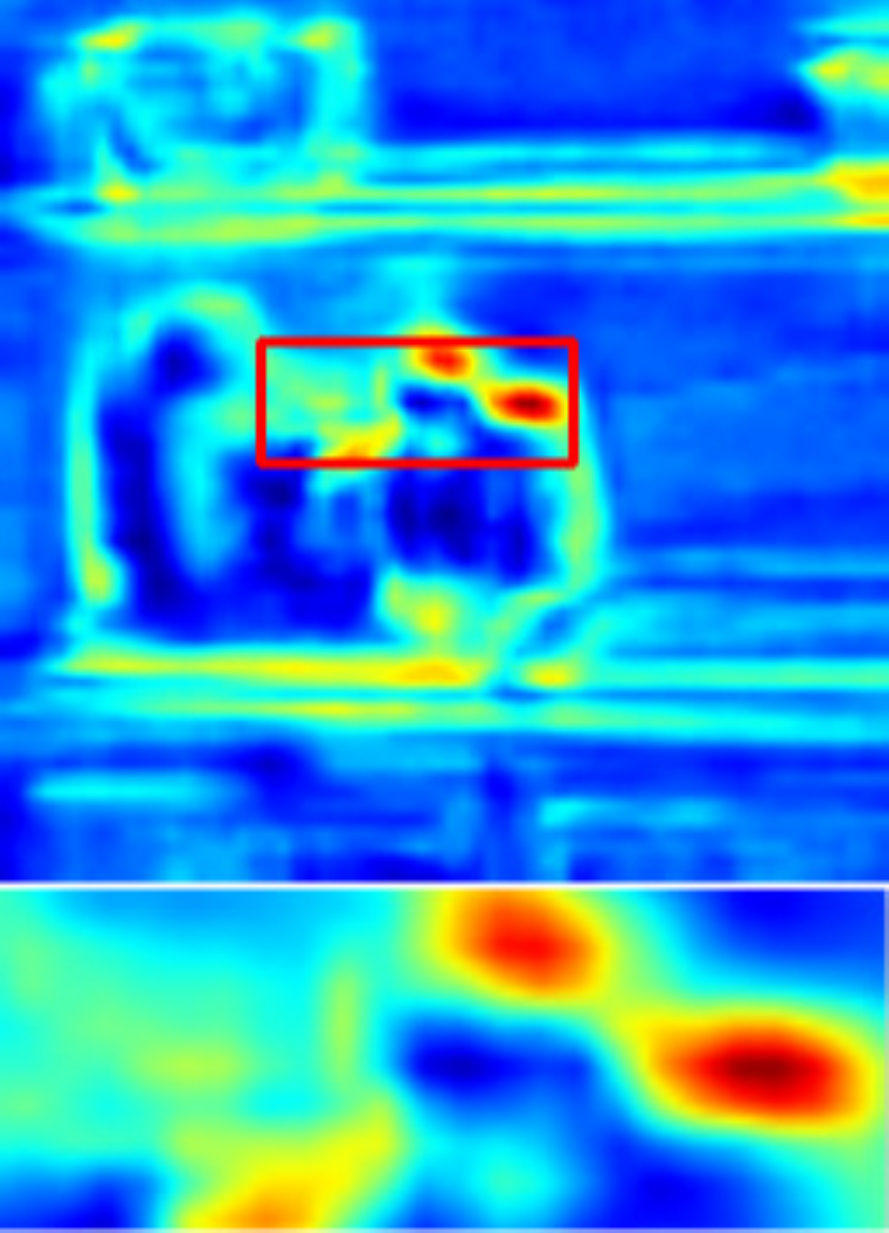}&\hspace{-5mm}
\includegraphics[width = 0.19\linewidth]{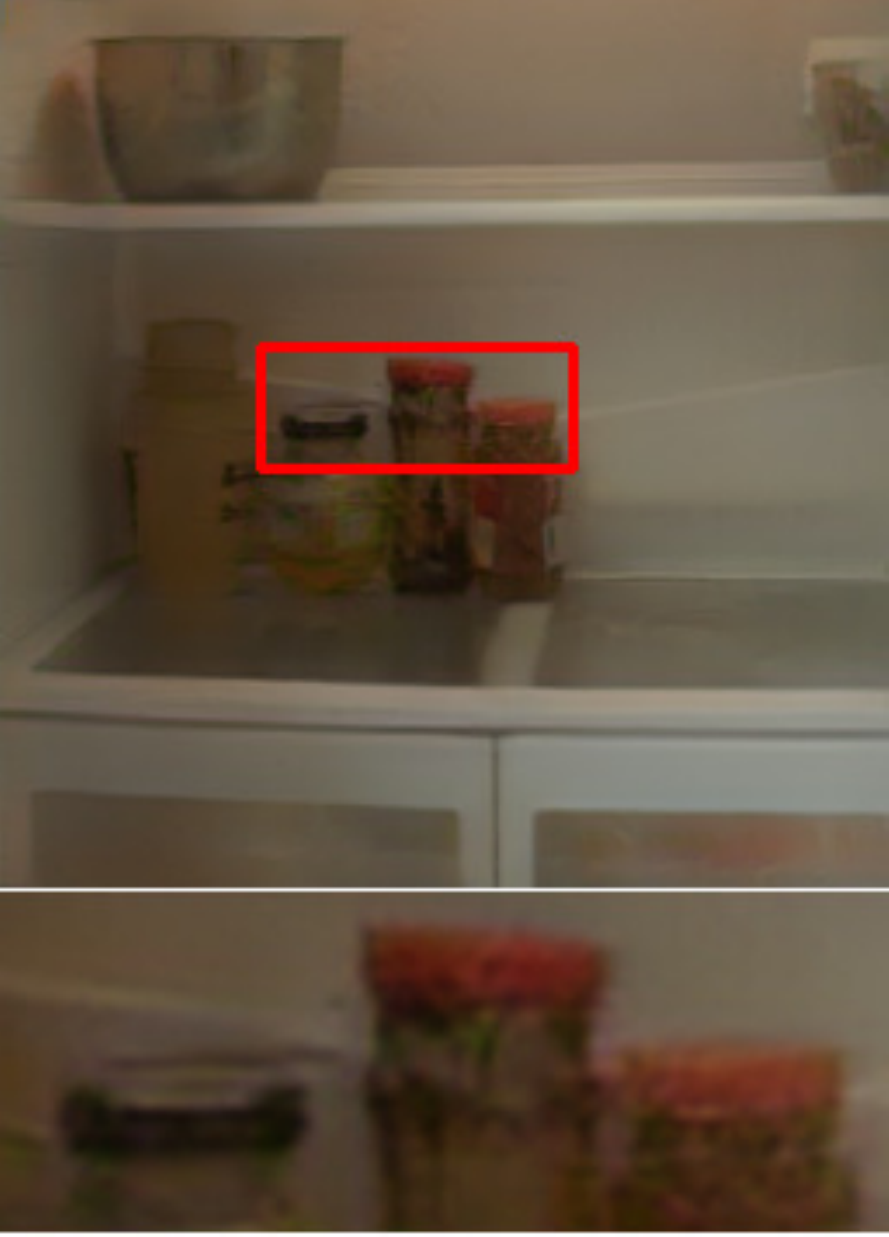}&\hspace{-5mm}
\includegraphics[width = 0.19\linewidth]{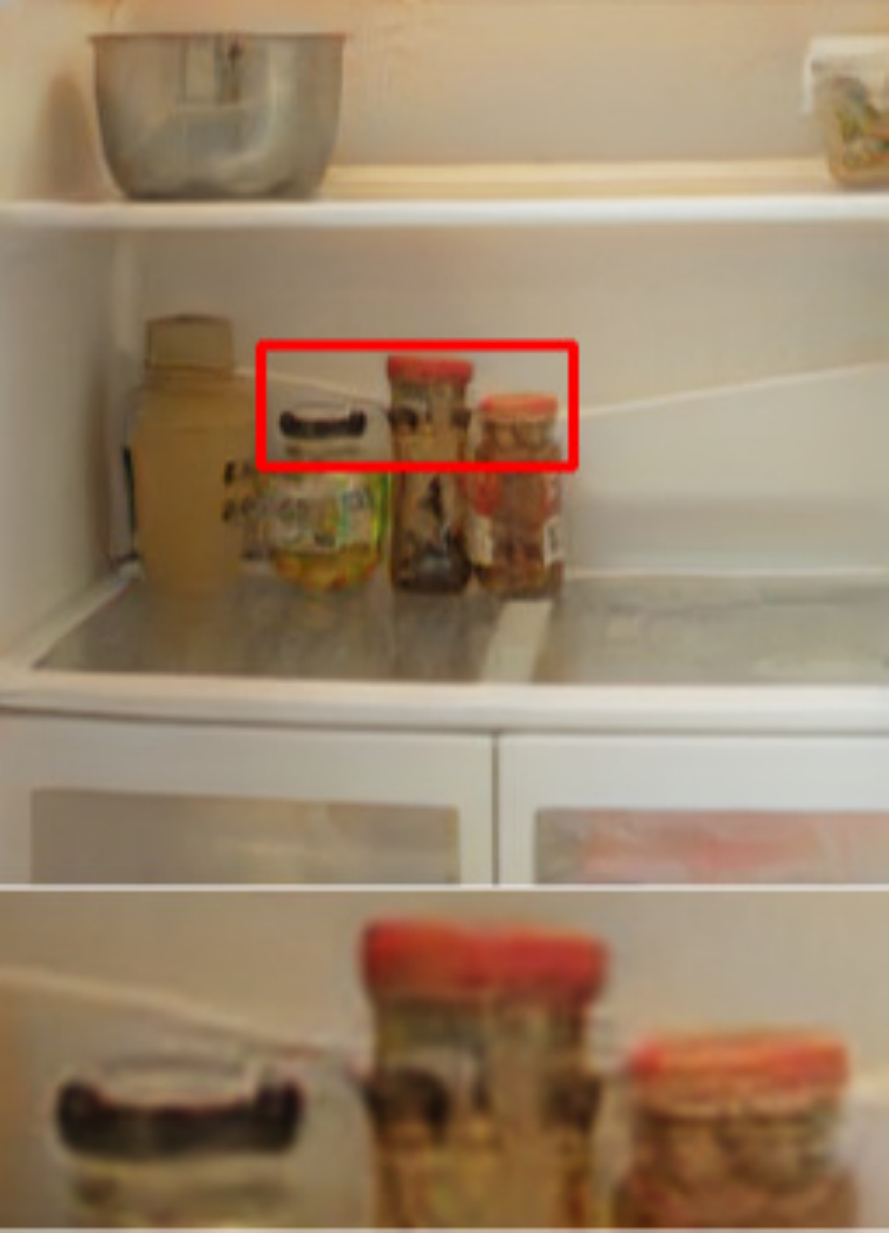}
\\
(a) &\hspace{-5mm} (b)  &\hspace{-5mm}  (c) &\hspace{-5mm} (d) &\hspace{-5mm} (e)
\\
\end{tabular}
\end{center}
\caption{Images enhanced by attention module, edge module and LEUGAN. (a) is the original image, (b) is without attention module, (c) is the attention map visualization, (d) is without edge module, (e) is our method. We can find that our results have little noise after the details are magnified, while most other images have higher noise.}
\label{fig:Ablation}
\end{figure}

\begin{table}[t]
\small
\begin{center}
\caption{Quantitative evaluation of low-light image enhancement algorithms. }
\label{table:1}
\begin{tabular}{cccc}
\hline\noalign{\smallskip}
Model & NIQE($\downarrow$)  & Vollath($\uparrow$) & PCA-based($\downarrow$)  \\
\noalign{\smallskip}
\hline
\noalign{\smallskip}
AMSR \cite{amsr} & 5.529 & 319.45 & 7.718  \\
BIMEF \cite{17} & 5.533 &672.37 & 5.896  \\
BPDHE \cite{bpdhe} & 5.163 & 2160.22 & 7.022  \\
Dong \cite{dong} & 5.669 & 1507.61 & 10.930  \\
LIME \cite{LIME} & 6.911 & {\color{blue}{\textbf{3248.44}}} & 13.660  \\
EnlightenGAN \cite{31}& 4.784 & 1760.56 & 7.562 \\
Retinex-Net \cite{62} &  5.808 & 872.54 & 10.268 \\
NPE \cite{npe} & 5.678 & 1649.80 & 13.218  \\
MBLLEN \cite{MBLLEN}& 4.670 & 2408.47 & 12.365 \\
GLADNet \cite{GLAD}& 6.067  & 2380.35 &5.542 \\
CycleGAN \cite{cyclegan}& {\bf 4.263} & 2235.47 &{\bf 1.542} \\
SRIE \cite{7} & 5.633 & 608.43 & 5.140 \\
LEUGAN & {\color{blue}{\textbf{4.216}}} & {\bf 2420.34} & {\color{blue}{\textbf{0.616}}}  \\
\hline

\end{tabular}
\end{center}
\end{table}

\subsection{Ablation Study}
We conduct ablation studies on the dataset to evaluate the effectiveness of different components in the method from visual effects and quantitative analysis based on the low-light dataset. Fig.~\ref{fig:Ablation} shows the ablation results of our different modules.

\textbf{Attention module.} Fig.~\ref{fig:Ablation}~(b) illustrates an image that is enhanced via a network without attention module. Clearly, it is dull in color while edges are blurred.
With the help of the attention maps shown in Fig.~\ref{fig:Ablation}(c), the generator could capture the global structure (e.g., the brightness of the image) as well as pay more attention to some local important areas so that many previously lost image details can be recovered, and the colors become richer and more natural.

\textbf{Edge module.} Fig.~\ref{fig:Ablation}(d) shows the result with edge module, and we can see that our result shown in Fig.~\ref{fig:Ablation}(e) generates high-quality images with sharp edges and smooth surfaces under the guidance of the edge module.

\section{Conclusion}

In this paper, we proposed a low-light image enhancement network using unsupervised generative attentional networks with designed attention and edge modules period, and these two modules help improve the image quality in terms of restoring color and sharpening edges.
We test our method on the LOL dataset and compare it with some state-of-the-art methods. Results show that our method outperform the others approaches in terms of image clarity and amount of noise.



%
%
\bibliographystyle{IEEEtran}
\bibliography{egbib}
\end{document}